\documentclass[apjl,twocolumn]{emulateapj_mod}
\usepackage{epsfig,apjfonts,mathptmx}

\def\gtsima{$\; \buildrel > \over \sim \;$}
\def\ltsima{$\; \buildrel < \over \sim \;$}
\def\prosima{$\; \buildrel \propto \over \sim \;$}
\def\gsim{\lower.5ex\hbox{\gtsima}}
\def\lsim{\lower.5ex\hbox{\ltsima}}
\def\simgt{\lower.5ex\hbox{\gtsima}}
\def\simlt{\lower.5ex\hbox{\ltsima}}
\def\simpr{\lower.5ex\hbox{\prosima}}

\def\h1{$h^{-1}$}
\def\eeq{\end{equation}}
\def\beq{\begin{equation}}
\def\24mu{24\,$\mu{\rm m}$}
\def\70mu{70\,$\mu{\rm m}$}
\def\8mu{8\,$\mu{\rm m}$}

\shorttitle{Low efficient star formation in $z\sim1.5$ ULIRGs}
\shortauthors{E. Daddi et al.}

\begin{document}

\title{Vigorous star formation with low efficiency in massive disk galaxies at $\lowercase{z}=1.5$\footnote{Based on observations carried out with the IRAM Plateau de Bure Interferometer. IRAM is supported by INSU/CNRS (France), MPG (Germany) and IGN (Spain).}}

   \author{E. Daddi\altaffilmark{1},
           H. Dannerbauer\altaffilmark{2},
           D. Elbaz\altaffilmark{1},
           M. Dickinson\altaffilmark{3},
           G. Morrison\altaffilmark{4,5},
           D. Stern\altaffilmark{6},
           S. Ravindranath\altaffilmark{7}
           }

\altaffiltext{1}{Laboratoire AIM, CEA/DSM - CNRS - Universit\'e Paris Diderot,
    DAPNIA/Service d'Astrophysique, CEA Saclay, Orme des Merisiers,  91191 Gif-sur-Yvette Cedex, France {\em edaddi@cea.fr}}
\altaffiltext{2}{MPIA, K\"onigstuhl 17, D-69117 Heidelberg, Germany}
\altaffiltext{3}{NOAO,  950 N. Cherry Ave., Tucson, AZ, 85719}
\altaffiltext{4}{Institute for Astronomy, University of Hawaii, Honolulu, HI 96822, USA}
\altaffiltext{5}{Canada-France-Hawaii Telescope, Kamuela, HI, 96743, USA}
\altaffiltext{6}{Jet Propulsion Laboratory, Caltech, Pasadena, CA 91109}
\altaffiltext{7}{IUCAA, Pune University Campus, Pune 411007, Maharashtra, India}

\begin{abstract}
We present the first detection of molecular
gas cooling CO emission lines from ordinary massive galaxies at $z=1.5$.
Two sources were observed with the IRAM Plateau de Bure Interferometer,
selected to lie in the mass-star formation rate
correlation at their redshift, thus being representative of massive high-$z$
galaxies. Both sources were detected with high confidence, yielding
$L'_{\rm CO}\sim2\times10^{10}$~K~km~s$^{-1}$~pc$^2$.
For one of the sources we find evidence for velocity shear,
implying CO sizes of $\sim10$~kpc.
With an infrared luminosity of $L_{\rm FIR}\sim10^{12}L_\odot$, these 
disk-like galaxies are borderline ULIRGs but with
star formation efficiency similar to that of local spirals, 
and an order of magnitude
lower than that in submm galaxies. 
This suggests a CO to total gas conversion factor similar to local 
spirals, gas consumption timescales approaching 1~Gyr or longer and
molecular gas masses reaching $\sim10^{11}M_\odot$, comparable to 
or larger than the
estimated stellar masses. 
These results support a major role of {\em in situ} gas
consumption over cosmological timescales and with relatively low
star formation efficiency, analogous to that of local spiral disks,
for the formation of today's most massive galaxies and their central black
holes. Given the high space density of similar galaxies, 
$\sim10^{-4}$~Mpc$^{-3}$, this implies a widespread presence of gas rich 
galaxies in the early Universe, many of which might be
within reach of detailed investigations of current and planned
facilities.
\end{abstract}

\section{Introduction}

Galaxies in the distant Universe were forming stars with much higher rates
than today (Madau et al 1996; Elbaz et al 2002; Le Floc'h et al. 2005).
For galaxies of a fixed stellar mass, e.g. of the order of
$M^*\sim10^{10-11}M_\odot$, 
the typical star formation rate (SFR) has been
steadily increasing with respect to local values, 
up to $\sim30$ times higher at 
$z=2$ (Daddi et al. 2007a; D07a hereafter).
This increase can be accounted
for either by postulating a higher gas content or a higher star formation
efficiency, or both. Galaxy mergers and interactions, expected to be much
more common in the distant Universe than today, can increase the
density of gas in galaxies, and thus help transform gas into stars
more rapidly and with a higher star formation
efficiency. 
The relative importance for massive galaxy formation and assembly 
of the hierarchical merging 
of small sub galactic units, versus the in situ formation of stars through gas 
consumption
is largely unknown observationally. 

The 
molecular gas content of galaxies is best probed
through the detection of redshifted
cooling emission lines of CO.
Combining observations for local and distant 
galaxies, all the way from spirals
to (ultra) luminous infrared galaxies  (LIRGs, $L_{\rm FIR}\simgt10^{11}L_\odot$, 
and ULIRGs, 
$L_{\rm FIR}\simgt10^{12}L_\odot$), a nonlinear correlation is observed
(e.g., Solomon \& van den Bout 2005)
between CO luminosity 
(proportional to molecular gas content) and FIR luminosity (proportional
to the SFR), 
with more luminous sources having higher FIR to CO luminosities,
i.e. higher efficiencies
than spiral disks  in converting
gas into stars, likely due to their higher gas densities
(Tacconi et al. 2006; Downes \& Solomon 1998).
Mapping the physics of molecular gas in representative
galaxies at high redshifts  would be a major
step forward into understanding the processes regulating galaxy formation,
and is a key target of future astrophysical facilities such as ALMA.

However, if the local correlations hold 
at high redshifts, current techniques and
sensitivities allow us to detect CO lines, and thus study the molecular
gas content, only for extreme objects such as bright ($S_{850\mu m}>5$~mJy)
submm-selected 
galaxies (Genzel et al. 2004; Neri et al. 2003; Greve et al. 2005).
(SMGs), quasars (Walter et al. 2003; Maiolino et al. 2005), 
or in the case of strong
gravitational lensing. 
Indirect evidence has been mounting very recently that this might not
be the general case. Vigorous starbursts at high redshift appear to be 
the ordinary star formation mode in massive galaxies. The existence
of tight correlations between SFRs and galaxy 
masses
(Elbaz et al. 2007; Noeske et al. 2007; D07a)
and the predominance of ULIRGs inside mass limited 
samples of $M\simgt5\times10^{10}M_\odot$ galaxies
(Daddi et al 2005; D07a; Caputi et al 2006)
suggests that vigorous starbursts were typically
long lasting among ordinary massive galaxies in the distant Universe, 
which therefore  might be hosting large amounts of gas.

\begin{figure}
\centering
\includegraphics[width=8.0cm,angle=0]{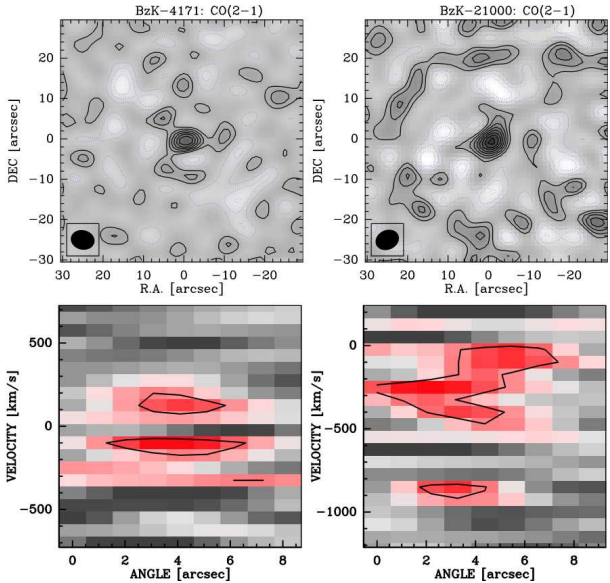}   
\caption{Top panels:
CO [2-1] emission line detections in two ordinary, massive high-$z$
galaxies. We show (uncleaned) velocity 
averaged spatial maps of CO emission, from
-350~km~s$^{-1}$ to 425~km~s$^{-1}$ ($BzK-4171$, left) and between  -575~km~s$^{-1}$ and
-25~km~s$^{-1}$ and from -950~km~s$^{-1}$ and -800~km~s$^{-1}$ ($BzK-21000$, right).
The contours are from -3 to
$+6\sigma$,
in steps of $1\sigma$ (0.1~mJy/beam). The field of view is 1$'$.
The size of the beam is shown in the lower-left.
Bottom panels: position-velocity maps, extracted along the major axis direction
measured on the HST+ACS UV rest frame imaging. The spatial field of view is 
$\sim9''$. Red/white colors are for positive pixels.
Contours are at the level of 1~mJy/beam.
}
\label{Pic}
\end{figure}

\section{CO observations of B$\lowercase{z}$K galaxies}

Motivated by this evidence, we observed 
two representative massive 
galaxies at high redshift
with the IRAM Plateau de Bure 
Interferometer, 
searching for the redshifted emission lines from the CO [2-1] transition
at rest frame 230.539~GHz, 
with the aim of constraining their molecular
gas content.
The two galaxies, $BzK$-4171 (J123626.53+620835.3, VLA radio frame) 
and $BzK$-21000 (J123710.60+622234.6), at redshift $z=1.465$ and 
$z=1.522$, respectively (measured from Keck DEIMOS spectroscopy), 
were selected in the Great Observatories Origins Deep
Survey Northern field (GOODS-N) using the $BzK$ technique 
(Daddi et al 2004b), which
provides
complete $K$-limited samples of normal, massive galaxies 
at $1.4<z<2.5$. 
SFRs of about 100-150 $M_\odot$~yr$^{-1}$
were derived by averaging estimates based on the VLA radio detections
of $40\pm5\mu$Jy and $49\pm7\mu$Jy (using the radio-$L_{\rm FIR}$ correlation,
Condon et al. 1992) and
UV estimates (correcting for dust reddening using the Calzetti et al. 2000 
law), which agree within 0.2~dex or better. For $BzK-4171$
the 24$\mu$m estimate (based on the Chary and Elbaz 2001 luminosity dependent templates) also agrees within this factor, while for $BzK-21000$ the
24$\mu$m estimate of SFR is 2-2.5 larger than the radio and UV ones, mild
evidence for possible mid-IR excess usually linked to obscured AGN
activity (Daddi et al. 2007b; D07b hereafter). 
Stellar masses of 0.5--0.8$\times10^{11}M_\odot$
were derived from optical and UV rest
frame photometry, following the method of Daddi et al. (2004b). SFRs and 
stellar masses are given here for the Chabrier (2003) stellar
initial mass function.
The two target galaxies
lie in the mass-SFR correlation established at their redshifts
(D07a), 
and are thus ordinary massive high-$z$ sources with
ongoing star formation.

\begin{figure}
\centering
\includegraphics[width=6.5cm]{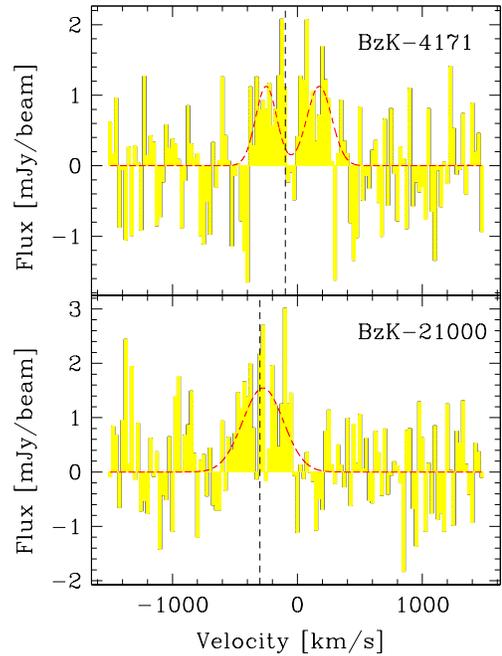}
\caption{
We show the CO spectra in velocity space, with respect
to the zero velocity offsets (dashed vertical lines)
defined from the optical redshift
measured
from the [OII] emission lines (we assume an intensity weighted wavelength of 3727.3\AA\ for
the doublet). 
Best fitting Gaussian profiles are shown for the emission of the 2
targets. 
}
\label{Spe}%
\end{figure}

The sources were observed
using newly refurbished receivers for
5.7 and 7.5 hours during April 24-25th 2007 for $BzK$-4171 and $BzK$-21000, 
respectively, during
stable and high quality conditions. A compact configuration-D with 6 antennas
was used.
Additional 7.5
hours of poorer quality weather data in configuration-D
and with 5 antennas only
were obtained for $BzK$-4171 on August 1st 2007,
reducing the noise by only 25\%. Standard data reduction
was performed using the GILDAS software.
Both sources observed
were detected in CO with high confidence (Fig.~\ref{Pic}, top panels), 
with a significance of 6.8$\sigma$ and 8.9$\sigma$.
Fluxes are
$0.60\pm0.09$~Jy~km~s$^{-1}$ and $0.85\pm0.10$~Jy~km~s$^{-1}$,
for $BzK$-4171 and $BzK$-21000, respectively.
Fig.~\ref{Spe} shows the CO spectra of the two $BzK$ galaxies.
The CO lines wavelengths are in agreement with the Keck 
redshifts, within 50~km~s$^{-1}$ or better. The CO positions agree with 
the VLA positions within $\approx0.5''$.
$BzK$-4171 is best fitted by a double horn profile, indicative
of a rotating disk and consistent with the UV morphology of a nearly edge
on disk. $BzK$-21000 shows
a narrower velocity profile with only a hint of double-horns. 
Fig.~\ref{Pic} (bottom panels) shows position-velocity diagram, derived 
using a pseudo-slit oriented along the major axis defined in the HST+ACS
imaging. $BzK$-4171 is unresolved, while  $BzK$-21000 shows evidence
for shear, overall consistent with rotation in a disk. As a cross-check, 
we separately
fitted   velocity channels bluer and redder than the systemic velocities.
For $BzK$-4171 the positions of the blue and red horns are separated by
$0.5\pm0.5''$. For $BzK$-21000 we find $1.3\pm0.3''$ separation ($11\pm2.5$~kpc; $4.3\sigma$),
definying a position angle of $-10\pm17$~deg, in very good agreement with
the UV position angle, confirming that the CO emission is extended.

\begin{figure*}
\centering
\includegraphics[width=15.5cm]{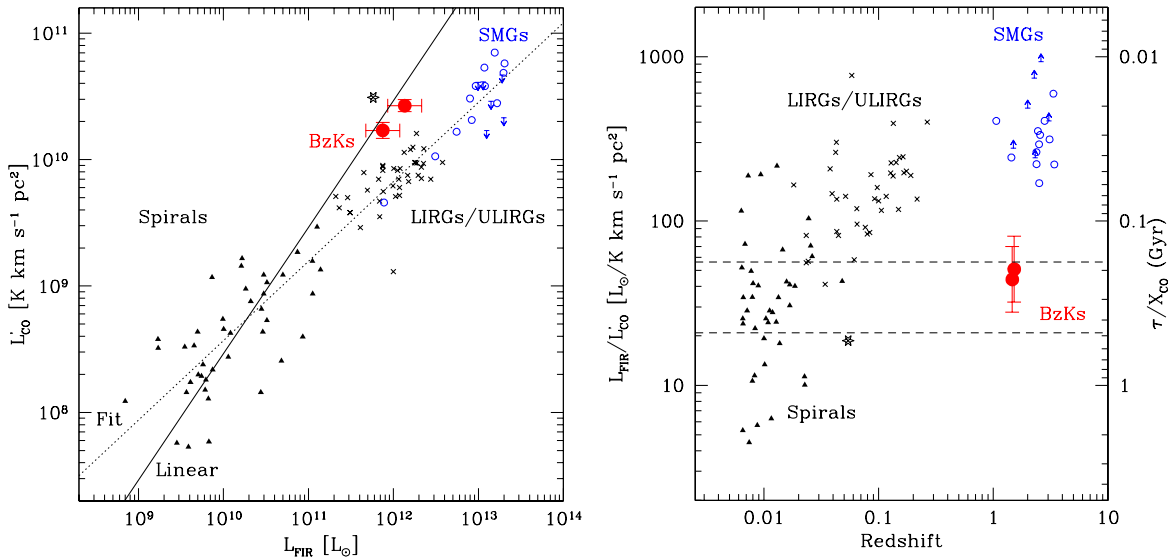}
\caption{
Comparison of molecular CO luminosity ($L'_{CO}$) to far-infrared luminosity
($L_{FIR}$, integrated between 40 and 500$\mu$m).
Quantities for the 2 $BzK$ galaxies presented
in this work are shown as large filled circles. 
The left panel shows
the comparison of the 2 luminosities. The dotted line is the best fit relation
to spirals (triangles), LIRG/ULIRGs (crosses) and distant SMGs (empty circles).
The solid line is for a linear $L'_{CO}$ to $L_{FIR}$ correlation
normalized to the ratio of local spirals.
The local LIRG with high $L'_{CO}$ content, behaving similarly to
distant $BzK$s, is VII~Zw~31 (plotted as an asterisk; Downes \& Solomon 1998), 
which interestingly
also has fairly cold 60/100$\mu$m color
ratios and larger CO size compared to other ULIRGs.
The right panel shows the $L_{FIR}$
to $L'_{CO}$ ratio as a function of redshift. The dashed horizontal lines
show the semi-interquartile range of local spirals.
}
\label{fig:XCO}
\end{figure*}

\section{Gas rich massive galaxies at $z\sim1.5$}

These results unveil a 
new and different galaxy formation mode for the most luminous
sources like LIRGs and ULIRGs.
In Fig.~\ref{fig:XCO} we compare CO to FIR luminosities of our $BzK$
galaxies to local and distant sources.
We adapted published luminosities
for local (Solomon et al. 1997; Yao et al 2003) and distant sources
(Greve et al. 2005)
to WMAP
cosmological parameters (Spergel et al. 2007).
We assume here roughly constant  brightness temperature
(see Braine \& Combes 1992)
and thus conversion
ratios among the different CO tracers to be of order unity, i.e. CO[1-0]$\sim$CO[2-1]$\sim$CO[3-2] (Wei{\ss} et al 2005).
Compared to local ULIRGs, these ordinary massive $z=2$ galaxies 
have similar bolometric luminosities but larger CO luminosities by
a factor of 3--5. On the other hand, SMGs have estimated
$L_{\rm FIR}\simgt10^{13}L_\odot$
and therefore (Kennicutt et al. 1998)
$SFR\simgt1000M_\odot$~yr$^{-1}$, one order of magnitude
higher than our two targets, but 
the observed CO luminosities are comparable.
The two $BzK$ galaxies are outliers of the $L_{\rm FIR}$ versus $L_{\rm CO}$
correlation defined by local spirals, LIRGs/ULIRGs and distant SMGs (Fig.~\ref{fig:XCO}).
Instead, they lie close to a linear $L_{\rm FIR}$ to $L_{\rm CO}$ relation normalized
to local spirals. 
The observed ratio
in the two massive $BzK$ galaxies is an order of magnitude lower  than
that of SMGs at similar redshifts, implying much lower star formation
efficiency. This is consistent with the fact that
SMGs are outliers of the mass-SFR correlation (D07a), probably due
to the higher efficiency in forming stars, for a similar stellar mass and
CO luminosity.
The $L_{\rm FIR}$ to $L_{\rm CO}$ ratios of the two $BzK$ galaxies are within
the semi-interquartile range of values 
from local spirals (Fig.~\ref{fig:XCO}, right panel). 
No object with similar $L_{\rm FIR}$ to $L_{\rm CO}$ 
ratio and large $L_{\rm FIR}$
was previously known in the distant Universe (Fig.~\ref{fig:XCO}). 
This implies that these distant BzK galaxies behave as 
{\em scaled up} versions of local spirals, with proportionally higher
SFR and molecular gas content for a fixed
galaxy stellar mass.

We use a conversion rate (Kennicutt et al. 1998)
$L_{\rm FIR}/L_\odot=0.8\times10^{10} SFR$~[$M_\odot$~yr$^{-1}$], adapted to
a Chabrier (2003) stellar initial mass function and in the assumption,
appropriate for these galaxies, that $L_{\rm FIR}$
is about  80\% of the bolometric luminosity. In general, we can write
$L'_{\rm CO}= M_{\rm H2}/X_{\rm CO}$, where $M_{H2}$ is the total molecular
gas mass (including He) and the conversion factor (Downes \& Solomon 1998;
Solomon \& van den Bout 2005) is
$X_{\rm CO}=4.6$~$M_\odot$~(K~km~s$^{-1}$~pc$^2$)$^{-1}$ for the Milky Way and as
low as $X_{\rm CO}=0.8$~$M_\odot$~(K~km~s$^{-1}$~pc$^2$)$^{-1}$
for  ULIRGs and likely
for SMGs.
The high detected CO luminosities imply that giant molecular 
gas reservoirs are present, of order of $2\times10^{10}M_\odot$,
for a ULIRG-like $X_{\rm CO}\sim1$, comparable to SMGs and larger than in 
local ULIRGs.
The typical gas consumption timescales (Fig.~\ref{fig:XCO}) are about 2--4$\times10^7$~yr for SMGs, a factor of $\approx2$
larger for local ULIRGs and reach 2--3$\times10^8$~yr
for the massive $BzK$ galaxies at $z=1.5$, in the case 
of $X_{\rm CO}\sim1$. 
The SFR could last for even longer than 
that, as further gas accretion will be substantial during 
such long timescales. Hydrodynamical simulations from Keres et al. (2005)
predict gas accretion rates of about 10-70 $M_\odot$~yr$^{-1}$ for similarly
massive galaxies at $1.4<z<2.5$ (i.e., for dark matter halo masses
of $\sim10^{12-13}M_\odot$; Hayashi et al 2007). This is
the same order of magnitude of the estimated SFR in massive $z\sim2$ 
galaxies.
These are likely lower limits to the SFR duration and gas masses
as we are entirely neglecting atomic gas. 
As the star formation efficiencies are consistent with those of local 
spirals, in analogy to the latter it might be that $X_{\rm CO}>1$ or even
that the Milky Way
conversion factor applies to these massive $BzK$
sources.

This is further supported by the 
evidence that, unlike local ULIRGs (Downes \& Solomon 1998) and distant 
SMGs (Tacconi et al 2006), 
these two $BzK$ galaxies, like $z>1.4$ 
massive star forming galaxies (Daddi et al 2004a; Bouche et al 2007) in general,
are not compact merger-induced starbursts. From detailed  S\'ersic 
fitting of their UV rest frame light profiles from Hubble Space
Telescope imaging data obtained with Advanced Camera for Surveys (Giavalisco
et al. 2004),
we derive  half-light radii of 4.4 and 5.5~kpc (proper size, or about 0.5$''$, 
accurate to better than 5\%). Both galaxies have
disk-like (exponential) morphology with S\'ersic index $n=1.3$ and 0.6
(accurate to better than 20\%). 
BzK-4171 appears indeed as a nearly edge on disk,
consistent with the S\'ersic fitting. BzK-21000 has a clumpy structure, 
still consistent with a disk but also with the possible presence of
minor merging events.
The UV rest frame sizes are likely reasonably 
accurate measurements of the starbursts sizes, because for these sources
reliable SFRs measurements can be obtained from the UV (Daddi et al 2005; D07a;
Dannerbauer et al 2006; Reddy et al 2005), i.e. their
UV emission is not entirely opaque, while
local ULIRGs and distant SMGs are optically thick in the UV 
(Champan et al 2005; Goldader et al 2002). This is confirmed by the evidence
for extended CO emission, at least for BzK-21000.
The inferred sizes are a factor of 2-3 larger than distant 
SMGs  (Tacconi et al 2006) and 10 times larger than local ULIRGs
(Downes \& Solomon 1998).
This converts into 1--2 order 
of magnitudes lower gas densities 
and justifies the relatively modest star formation efficiencies.
Independent estimates of the molecular gas masses
and thus SFR timescales, can also be obtained.
First, if we apply the Schmidt-Kennicutt law 
(using its latest derivation from Bouche et al 2007, that includes
also distant SMGs), using
the UV size for the typical gas and SFR size, 
we estimate $M_{gas}\sim10^{11}M_\odot$ for each galaxy.
Second, from the CO velocity widths and (UV and/or CO) size
measurements of the galaxies
we estimate dynamical masses of about 1--1.5$\times10^{11}M_\odot$, within
the half light radii, including gas, stars and dark matter. 
Given the estimated total stellar masses of 
$\sim0.5$--0.8$\times10^{11}M_\odot$, roughly 
half of which presumably lying within the half-light radii, 
it looks plausible that $\approx10^{11}M_\odot$ of 
gas can be present.
We conclude that it
would take up to 1~Gyr  to consume all of the gas in these
typical, massive high-redshift galaxies, assuming the star-formation
rate remained roughly constant.  Clearly, the star formation episodes
could last longer since additional gas is accreted over such long
timescales.

\section{Discussion}

In order to evaluate the implications of these results for massive galaxy 
formation,  recall that the two observed galaxies
are representative of massive galaxies at high
redshifts. They were selected  for CO observations simply due to the
availability of
known spectroscopic redshifts. 
The two galaxies lie on the 
SFR to stellar mass correlation inferred at $1.4<z<2.5$, and their colors and morphologies are fully typical of massive $z\sim2$ galaxies.
Star forming galaxies in that redshift range with similar
stellar masses are common
in the distant Universe, with space densities of order 
of $10^{-4}$~Mpc$^{-3}$, quite a bit larger than the
$2-6\times10^{-6}$~Mpc$^{-3}$ observed
for SMGs by Chapman et al. (2003). Clearly, these results
point to the existence of a much larger population of gas rich galaxies 
in the distant Universe than previously thought.

We notice that the far-IR luminosity estimates for our objects are based
on the UV, radio and mid-IR luminosities, and are solid given that
self-consistent measurements of SFR activity have been shown to be obtainable
for these kind of objects at high redshift (D07a). 
It is very unlikely
that we are strongly underestimating their star formation activity
(which would lead us to overestimate the duration timescales).
Instead, the $L_{\rm FIR}$ for SMGs from Greve et al. (2005) is based
only on the $850\mu$m flux and is likely more uncertain. Using radio and
Spitzer, Pope et al. (2006) suggest lower $L_{\rm FIR}$ for SMGs.
This could imply that SMGs are closer in their SFR modes to
massive $BzK$ galaxies, but would change little about our
conclusion because it is the latter which represent the dominating star
formation mode at high redshift.

The large molecular gas fraction in these galaxies, possibly up to
50\% or more, would likely give rise to large scale 
instabilities, with the formation of clumps (e.g., Immeli et al. 2004;
Bournaud, Elmegreen \& 
Elmegreen 2007). This could explain the clumpy structure of these
two galaxies, and in general of
a large fraction of the high-$z$ massive galaxies 
(e.g., Daddi et al. 2004a), and the high velocity dispersions observed in
many of the high-$z$ massive disk galaxies (e.g., Genzel et al. 2006; 
Forster-Schreiber et al. 2006).

The long duration timescale for star formation activity implies, as a corollary,
long duration activity for the galactic nuclei. 
In fact, it has been recently
unveiled by D07b 
that up to 20-50\% of massive galaxies in $1.4<z<2.5$
likely contain strongly obscured but relatively luminous AGNs. 
Some of the large  gas reservoir is likely slowly funneled, in
some way,
into the central black hole with similar timescales to star formation, 
at a rate roughly consistent with what is required to assemble the  
correlation between stellar and black hole masses 
as we observe today (D07b).
Our results underline the major role
of gas consumption over long timescales and with low efficiencies,
as opposed to rapid and strong merger driven bursts, as a major growth 
mode for both stellar mass and black holes in the distant Universe.

\smallskip
We thank Roberto Neri and Arancha Castro-Carrizo for assistence with
taking and reducing IRAM Plateau de Bure Interferometer data.
We  acknowledge very useful discussions, at various stages of this work,
with Kartik Sheth, Frederic Bournaud, Pierre-Alain Duc and Fabian Walter.
Comments by an anonymous referee improved the paper.
We acknowledge the use of GILDAS software (http://www.iram.fr/IRAMFR/GILDAS).
Support for this work, part of the Spitzer Space Telescope
Legacy Science Program, was provided by NASA, Contract Number 1224666
issued by the JPL, Caltech, under NASA contract 1407.

\end{document}